\documentclass[11pt,hyper,letterpaper]{JHEP3}
\usepackage{graphicx,amssymb,amsmath,amsfonts,epstopdf,bm,cite}

\def\href#1#2{#2}

\def\beq{\begin{equation}}
\def\eeq{\end{equation}}

\def\N{{\mathcal{N}}}

\def\<{\langle}

\def\>{\rangle}
\def\({\left (}
\def\){\right )}
\def\[{\left[}
\def\]{\right]}

\def\beq{\begin{equation}}
\def\eeq{\end{equation}}

\def\jt{\langle J^t \rangle}

\newcommand{\bea}{\begin{eqnarray}}
\newcommand{\eea}{\end{eqnarray}}

\def\ra{\rightarrow}

\def\k{\kappa}             

\def\o{\omega}  

\title{ \LARGE Zero Sound in Strange Metallic Holography}
\author{Carlos Hoyos,$^2$\footnotemark[1]\, Andy O'Bannon,$^1$\footnotemark[2]\, and Jackson M. S. Wu$^3$\footnotemark[3]
\\
\\
$^1$Department of Physics, University of Washington \\ Seattle, WA 98195-1560, United States
\\
\\
$^2$Max-Planck-Institut f\"{u}r Physik (Werner-Heisenberg-Institut) \\ F\"{o}hringer Ring 6, 80805 M\"{u}nchen, Germany
\\
\\
$^3$Albert Einstein Center for Fundamental Physics \\
Institute for Theoretical Physics, University of Bern \\ Sidlerstrasse 5, 3012 Bern, Switzerland}

\footnotetext[1]{E-mail address: \email{choyos@phys.washington.edu}}
\footnotetext[2]{E-mail address: \email{ahob@mppmu.mpg.de}}
\footnotetext[3]{E-mail address: \email{jbnwu@itp.unibe.ch}}

\abstract{One way to model the strange metal phase of certain materials is via a holographic description in terms of probe D-branes in a Lifshitz spacetime, characterised by a dynamical exponent $z$. The background geometry is dual to a strongly-interacting quantum critical theory while the probe D-branes are dual to a finite density of charge carriers that can exhibit the characteristic properties of strange metals. We compute holographically the low-frequency and low-momentum form of the charge density and current retarded Green's functions in these systems for massless charge carriers. The results reveal a quasi-particle excitation when $z < 2$, which in analogy with Landau Fermi liquids we call zero sound. The real part of the dispersion relation depends on momentum $k$ linearly, while the imaginary part goes as $k^{2/z}$. When $z\geq2$ the zero sound is not a well-defined quasi-particle. We also compute the frequency-dependent conductivity in arbitrary spacetime dimensions. Using that as a measure of the charge current spectral function, we find that the zero sound appears only when the spectral function consists of a single delta function at zero frequency.}

\keywords{AdS/CFT correspondence, Gauge/gravity correspondence}
\preprint{MPP-2010-76}
\begin{document}

\section{Introduction and Summary}
\label{intro}

Gauge-gravity duality~\cite{Maldacena:1997re,Gubser:1998bc,Witten:1998qj} provides a novel way to compute observables in strongly-coupled, scale-invariant systems at finite density, and thus may be useful for studying condensed matter systems near quantum critical points, such as the ``strange metal'' phase of some heavy fermion compounds and possibly also of some high-$T_c$ materials. In particular, gauge-gravity duality may be able to reproduce the simple properties of strange metals, such as electrical resistivity scaling linearly with temperature $T$, which cannot be derived in a Fermi liquid description, which assumes the appropriate degrees of freedom are weakly-interacting quasi-particles. Of course the ultimate goal is to understand why some strange metals have a high $T_c$.

Gauge-gravity duality is holographic: it equates a weakly-coupled theory of gravity on some spacetime with a strongly-coupled field theory living on the boundary of that spacetime. The spacetime symmetries of the field theory are dual to the isometries of the bulk metric. Here we will focus on field theories invariant under the Lifshitz group, which includes scale transformations of the time coordinate $t$ and spatial coordinates $\vec{x}$ of the form
\beq
t \ra \lambda^z \, t \,, \qquad \vec{x} \ra \lambda \, \vec{x} \,,
\eeq
where $\lambda$ is real and positive and $z$ is called the ``dynamical exponent.'' The dual gravity theory then lives in Lifshitz spacetime, with metric \cite{Kachru:2008yh}
\beq
ds^2 = g_{\mu\nu} \, dx^{\mu} dx^{\nu} = \frac{dr^2}{r^2} - \frac{dt^2}{r^{2z}} + \frac{d\vec{x}^2}{r^2} \,,
\eeq
where $r$ is the holographic radial coordinate: $r \rightarrow \infty$ corresponds to the infrared (IR) of the field theory, while the near-boundary region $r \rightarrow 0$ corresponds to the ultraviolet (UV). Notice that when $z=1$, the metric becomes that of anti-de Sitter space (AdS), and the dual field theory symmetry is enhanced to the relativistic conformal group.

One proposal for holographic model building of strange metals is to use probe D-branes in a Lifshitz spacetime~\cite{Hartnoll:2009ns}. In this approach, an appropriate configuration of fields on the probe D-brane represents a finite density of massive charge carriers~\cite{Kobayashi:2006sb,Karch:2007pd,Karch:2007br}, while the background spacetime represents a strongly-coupled neutral quantum critical theory. Holographic calculations then reveal that such systems can indeed reproduce the simple properties of strange metals, including electrical resistivity linear in temperature~\cite{Hartnoll:2009ns}. However, ingredients beyond Lifshitz spacetime, for example a nontrivial dilaton, are needed to reproduce all the measured strange metal properties~\cite{Hartnoll:2009ns,Lee:2010ii}.

When $z=1$ and $T=0$, holographic calculations also reveal the existence of a propagating quasi-particle excitation producing a pole in the charge density and current retarded two point functions, which we denote $G_R^{tt}(\omega,k)$ and $G_R^{xx}(\omega,k)$, where $\omega$ is frequency and $k$ is momentum~\cite{Karch:2008fa,Kulaxizi:2008kv,Kim:2008bv}. In a weakly-coupled Fermi liquid, fluctuations in the shape of the Fermi surface produce a collective excitation at zero temperature, Landau's so-called ``zero sound.'' The dispersion relation of the mode in these holographic systems was identical in form to that of Landau's zero sound, with real part linear in $k$ and imaginary part going as $k^2$, hence the mode was dubbed zero sound by analogy. We will call this mode ``holographic zero sound.'' The physical origin of this mode remains mysterious since it appears in systems with both fermions and scalars---systems that do not fit neatly into either Fermi liquid or quantum Bose liquid theory. Indeed, given the existence of the holographic zero sound, as well as unusual scaling of the heat capacity with temperature, these holographic systems may be new kinds of (strongly-coupled) quantum liquids~\cite{Karch:2008fa}.

In this paper we explore the fate of holographic zero sound in these systems when $z > 1$. To do so, we compute holographically -- and completely analytically, without numerics -- the low-frequency and low-momentum forms of $G_R^{tt}(\omega,k)$ and $G_R^{xx}(\omega,k)$, for massless charge carriers. When $z\neq 2$ we find
\beq\label{eq:retsound}
G_R^{xx}(\omega,k)=\frac{\omega^2}{k^2} G_R^{tt}(\omega,k) \propto \frac{\omega^2}{ k^2 - \frac{1}{v^2} \, \omega^2- c \, \omega^{2/z+1}},
\eeq
where $v$ and $c$ are dimensionful constants that depend on the density.\footnote{When $z=2$ the last term in the denominator becomes $c \, \omega^2\log(\alpha \omega^2)$, with a scale $\alpha$ depending on the density.} We obtain the holographic zero sound dispersion relation by setting the denominator of $G^{tt}(\omega,k)$ to zero. Whether the mode is a quasi-particle depends on whether the $\omega^2$ or $\omega^{2/z+1}$ term is larger at low frequency. 
When $z<2$, the $\omega^2$ term dominates, and the mode remains a quasi-particle, although with a dispersion relation modified relative to the $z=1$ case: the real part remains linear, but the imaginary part goes as $k^{2/z}$. When $z>2$ the real and imaginary parts are of the same order, and hence the mode is no longer a sharply-defined quasi-particle.

We also compute the AC conductivity associated with the charge carriers, generalizing the results of ref.~\cite{Hartnoll:2009ns}, for two spatial field theory directions, to an arbitrary number of spatial directions. When $z<2$, the AC conductivity's dependence on $\omega$ is a pole formally identical to the high-frequency or ``collisionless'' limit of the standard Drude conductivity, $i \omega^{-1}$~\cite{Hartnoll:2009ns}, indicating that the charge current spectral function, which is a measure of the density of charged states, consists only of a delta-function term at $\omega=0$. When $z > 2$, the $\omega$-dependence is instead a power law of the form $\omega^{-2/z}$, indicating a nontrivial density of states at low frequency.

In short, we find that holographic zero sound appears in the \textit{absence} of low-frequency charged states. That alone is perhaps not surprising: a continuum of states can ``smear out'' a pole. Indeed, in interacting systems with a Fermi surface, the zero sound pole is most easily identified if the zero sound velocity is greater than the Fermi velocity, so that the pole is outside a continuum of particle-hole states.\footnote{For textbook reviews of zero sound in Fermi liquids, see for example section 1.7 of ref.~\cite{Pines:1966} or section 5.4 of ref.~\cite{Negele:1988vy}} The change in behavior of the dispersion relation at $z=2$ is the novel feature of the holographic systems.

This paper is organized as follows. We review the thermodynamics of these systems in section \ref{thermo}. We compute the Green's functions in section \ref{greens}, and use the results to study the holographic zero sound and AC conductivity in sections \ref{zero} and \ref{cond}, respectively. We conclude with some discussion and suggestions for future research in section \ref{discussion}.

\section{Probe Thermodynamics}
\label{thermo}

In this section we review the results of refs.~\cite{Karch:2007br,Karch:2008fa,Karch:2009eb,Benincasa:2009be,Lee:2010uy} for the thermodynamics of the charge carriers in these holographic systems.

In gauge-gravity duality, a theory of gravity is dual to some large-$N$, strongly-coupled non-Abelian gauge theory. $N_f$ probe D-branes are dual to $N_f$ fields in the fundamental representation of the gauge group(s) in the probe limit $N_f \ll N$~\cite{Karch:2002sh}. We will call these fields quarks or flavor fields, in analogy with Quantum Chromodynamics.

We will consider $N_f$ coincident D-branes, whose action is the non-Abelian Dirac-Born-Infeld (DBI) action, describing the dynamics of the worldvolume scalars and $U(N_f)$ gauge fields.\footnote{In string theory D-brane actions also include Wess-Zumino terms, describing the coupling of D-brane fields to background Ramond-Ramond fields. To keep the discussion general without committing to a specific string theory realization of Lifshitz spacetime~\cite{Hartnoll:2009ns}, we will omit Wess-Zumino terms.} The field theory then has a global $U(N_f)$ flavor symmetry. We produce a finite density of charge carriers by introducing a chemical potential for the diagonal $U(1) \subset U(N_f)$. The density operator $J^t$ is the time component of the conserved $U(1)$ current $J^{\mu}$, and is holographically dual to $A_t$, the time component of the $U(1)$ gauge field living on the D-branes. We will thus introduce in the bulk D-branes with only $A_t(r)$ present. The action then reduces to the Abelian DBI action.

We will only consider \textit{massless} flavor fields. A flavor field mass operator would be dual to a scalar field on the D-branes. We omit any such scalar from our calculations, although we comment on massive flavors in section~\ref{discussion}.

We assume our D-branes are extended along $q$ spatial dimensions of the Lifshitz spacetime.\footnote{If $q$ is less than the total number of spatial dimensions, then in the field theory the flavor fields propagate only along some $q$-dimensional defect.} The Abelian DBI action for our D-branes is then
\beq
\hat{S} = -N_f T_D V\!\!\int\!dr\,dt\,d^qx\,\sqrt{-\mathrm{det}\left[g_{ab} + (2\pi\alpha')F_{ab}\right]} \,,
\eeq
where $T_D$ is the tension, $g_{ab}$ the induced metric, and $F_{ab}$ the $U(1)$ field strength of the D-branes. The factor $V$ is the volume of any internal space that the D-brane may be wrapping, and $2\pi\alpha'$ is the inverse string tension, which will typically be present in an actual string theory system. Inserting our ansatz $A_t(r)$, performing the trivial integrations over $dt$ and $d^qx$, and dividing out the subsequent (infinite) volume factors, we obtain the action density $S$ (which for brevity we will call the action henceforth),
\beq
S = -\N\!\int\!dr\,g_{xx}^{q/2}\,\sqrt{|g_{tt}|g_{rr} - \tilde{A}_t^{'2}}\,, \qquad \N \equiv N_f T_D V \,,
\eeq
where a tilde denotes a factor of $2\pi\alpha'$, so $\tilde{A}_t \equiv (2\pi\alpha')A_t$, and the prime denotes $\partial_r$. The charge density in the field theory is
\beq
\jt = \frac{\delta S}{\delta A_t'} = \N g_{xx}^{q/2} \frac{(2\pi\alpha')\tilde{A}_t'}{\sqrt{|g_{tt}|g_{rr} - \tilde{A}_t^{'2}}} \,.
\eeq
Solving for $A_t'(r)$, we find
\beq
A_t'(r) = \frac{d}{2\pi\alpha'}\sqrt{\frac{|g_{tt}|g_{rr}}{g_{xx}^q +d^2}} \,,
\eeq
where $d\equiv\jt/\widetilde{\N}$. Inserting the solution for $A_t'(r)$ back into the action, we find
\beq
S = -\N\!\int\!dr\,|g_{tt}|^{1/2}g_{rr}^{1/2} \, \frac{g_{xx}^q}{\sqrt{g_{xx}^q + d^2}} \,.
\eeq

We obtain the field theory chemical potential and free energy by performing the integrals for $A_t'(r)$ and $S$. The chemical potential is
\beq
\label{eq:zeroTchemicalpotential}
\mu_0 = \int^{\infty}_0 dr\, A_t'(r) = \frac{d}{2\pi\alpha'}\int_0^{\infty}dr\,\frac{r^{-(z+1)}}{\sqrt{r^{-2q} + d^2}} = \frac{d^{z/q}}{4\pi\alpha'q}\frac{\Gamma\!\left(\frac{z}{2 q}\right)\Gamma\!\left(\frac{1}{2}-\frac{z}{2q}\right)}{\Gamma\left(\frac{1}{2}\right)}\,.
\eeq
The grand canonical (Gibbs) free energy density is
\beq
\label{eq:zeroTonshellaction}
\Omega_0 = -S = \N\!\int^{\infty}_0 dr\,\frac{r^{-(2q+z+1)}}{\sqrt{r^{-2q} + d^2}} =- \widetilde{\N}\frac{z\,d}{z+q}\mu_0 \,.
\eeq
These results are deceptively simple. We have actually removed various power-law divergences at the $r=0$ endpoint of the integrations via analytic continuation. To see the divergences explicitly, we expand the integrand of eq.~\eqref{eq:zeroTonshellaction} for small $r$,
\beq
\int dr\,\frac{r^{-(q+z+1)}}{\sqrt{1+ d^2 r^{2 q}}} = \int dr\,r^{-(q+z+1)}\left(1-\frac{1}{2} d^2 r^{2 q}+O\left(d^4 r^{ 4 q}\right) \right). 
\eeq
The first term gives a density-independent divergence for any $z$, coming simply from the volume of Lifshitz spacetime. As we increase $z$ to $q$ and beyond, new density-dependent UV divergences appear. These are proportional to even powers of $d$, and appear to be associated with operators $(J^t)^{2n}$ with $n\in\mathbb{N}$ that become relevant for larger values of $z$.\footnote{The scaling dimension of the current operator is $[J^t] = q$, while the volume element has dimension $[dt \, d^qx] = -z-q$, so as we increase $z$ to $q$ and beyond, the operator $J^t J^t$ goes from irrelevant to marginal (at $z=q$) to relevant. Similarly, when $z=3q$, $J^t J^t J^t J^t$ becomes marginal, and so on. Only even powers of $J^t$ can appear due to charge conjugation invariance $J^t \rightarrow -J^t$.} We could have cancelled all of these divergences by introducing a regulator by integrating only to a cutoff $r=\epsilon$ and then adding counterterms\footnote{We write the counterterms explicitly in section~\ref{greens}.} at $r=\epsilon$~\cite{deHaro:2000xn,Bianchi:2001kw,Karch:2005ms}, but these counterterms generally also introduce additional \textit{finite} density-dependent terms. Fixing these finite terms actually determines the thermodynamic potential, and thus the thermodynamic ensemble. We have implicitly fixed these terms via analytic continuation, and hence we have chosen an ensemble. A straightforward calculation reveals that for us
\beq
\langle J^t\rangle=-\frac{\delta \Omega_0}{\delta \mu_0}\,,
\eeq
which shows that we are working in the grand canonical ensemble. Notice that with massless flavor fields, any finite chemical potential will produce a finite density.

Analytic continuation cannot remove logarithmic divergences, which is indeed the case in eqs.~\eqref{eq:zeroTchemicalpotential} and~\eqref{eq:zeroTonshellaction} when $z = (2n-1)q$. These divergences occur because the operators $(J^t)^{2n}$ are classically marginal at these $z$ values~\cite{Hartnoll:2009ns}, and if added to the action, can produce a breaking of scale invariance through quantum corrections, similarly to what happens for multi-trace scalar operators~\cite{Witten:2001ua}.

The energy density, $\varepsilon_0$, is
\beq
\varepsilon_0 = \Omega_0 + \mu_0 \langle J^t \rangle = - \frac{q}{z} \Omega_0 \,.
\eeq
In the grand canonical ensemble and at zero temperature, the pressure $P_0$ is simply $P_0 = - \Omega_0 = (z/q)\varepsilon_0$, hence $(\partial P/\partial \varepsilon)_{\mu} = z/q$~\cite{Kim:2008bv,Lee:2010uy,Kulaxizi:2008jx}, a result fixed completely by Lifshitz scale invariance.\footnote{Holographic calculations of $(\partial P/\partial \varepsilon)_\mu$ using asymptotically Lifshitz black holes yield the same result (see for example ref.~\cite{Balasubramanian:2009rx}).} However, this quantity is only the speed of normal/first sound in the relativistic $z=1$ case. In general, the speed of normal/first sound is the derivative of the pressure with respect to the mass density, and the latter only coincides with the energy density in a relativistic system. In a system with $z>1$, a velocity $v$ must be dimensionful: given $[\omega] = z$ and $[k]=1$, we have $[v] = [\omega/k] = z-1$. Indeed, we will later find a holographic zero sound speed proportional to $d^{(z-1)/q}$.

To study thermodynamics at finite temperature, we need a black hole in the bulk whose Hawking temperature is dual to the field theory temperature~\cite{Witten:1998zw}. A number of asymptotically Lifshitz black holes have been studied recently -- for a sampling, see refs.~\cite{Taylor:2008tg,Danielsson:2009gi,Azeyanagi:2009pr,Mann:2009yx,Pang:2009ad,Bertoldi:2009dt,Balasubramanian:2009rx,Cai:2009ac,Ayonbeato:2010tm,Blaback:2010pp,Dehghani:2010gn}. To capture the basic qualitative physics without committing to a specific system, we will again follow ref.~\cite{Hartnoll:2009ns} and write a generic Lifshitz black hole metric
\beq
ds^2 = \frac{dr^2}{f(r) \, r^2} - \frac{f(r)}{r^{2z}} dt^2 + \frac{d\vec{x}^2}{r^2} \,.
\eeq
We do not need an exact form of $f(r)$. We only need to know that $f(r)$ has a simple zero at the horizon $r = r_H$, and that the temperature is fixed by the regularity of the Wick-rotated metric at the horizon,
\beq
\label{eq:Tdef}
f(r) \underset{r \ra r_H}{\simeq} c_0\left(1-\frac{r}{r_H}\right) \,, \qquad T = c_0\frac{r_H^{-z}}{4\pi} \,,
\eeq
where $c_0$ is a dimensionless constant that depends on the specific system.

The chemical potential at finite temperature is
\begin{align}
\mu= \int_0^{r_H}\!dr A_t' &= 
\mu_0-\frac{r_H^{-z}}{2\pi\alpha' \, z}\,
{_2F_1}\!\left(\frac{1}{2},\frac{z}{2q};1+\frac{z}{2q};-\frac{r_H^{-2 q}}{d^2}\right) \nonumber \\ 
&= \mu_0-\frac{r_H^{-z}}{2\pi\alpha' \, z}\left(1-\frac{z}{2(2q +z)}\frac{r_H^{-2 q}}{d^2} + O\left(\frac{r_H^{-4q}}{d^4}\right)\right) , \label{eq:smallTmu}
\end{align}
where the second line is a small-temperature expansion. The probe flavor contribution to the finite-temperature free energy is
\begin{align}
\Omega = -S & =\Omega_0-\N\frac{r_H^{-2q-z}}{(2q+z)d}\,
{_2F_1}\!\left(1+\frac{z}{2q},\frac{1}{2};2+\frac{z}{2q};-\frac{r_H^{-2q}}{d^2}\right) \nonumber \\
\label{eq:finiteTonshellaction}
&= -\widetilde{\N}\frac{z\,d}{z+q} \, \mu_0 + O\left(\frac{r_H^{-2q-z}}{d}\right) \nonumber \\
&= -\widetilde{\N}\frac{z\,d}{z+q}\left [ \mu + \frac{r_H^{-z}}{2\pi\alpha'z} \right] + O\left(\frac{r_H^{-2q-z}}{d}\right) \nonumber \\
&= -\widetilde{\N}\frac{z\,d}{z+q} \, \mu -\N \frac{4 \pi d \, T}{c_0(z+q)} + O\left(\frac{T^{2q/z+1}}{d}\right),
\end{align}
where in the second line we have used eq.~\eqref{eq:zeroTonshellaction}, in the third line eq.~\eqref{eq:smallTmu}, and in the fourth line eq.~\eqref{eq:Tdef}. The entropy density at zero temperature is proportional to the charge density~\cite{Karch:2008fa,Karch:2009eb,Lee:2010uy},
\beq
s_0 = -\left(\frac{\partial\Omega}{\partial T}\right)_{\mu,T=0} = \frac{4\pi \N}{c_0(z+q)} \, d.
\eeq
Indeed, in a string theory system, this entropy should be equivalent to the entropy of a single quark, represented in the bulk by a single string, times the density $d$ of quarks~\cite{Karch:2009eb}. A nonzero entropy at zero temperature, or equivalently a degeneracy of states, indicates a potential instability \cite{Charmousis:2010zz}, since typically any perturbation will break the degeneracy and the system will flow to a new, generically non-degenerate state. In other words, we may be working with states that are not the genuine ground state. The $z=1$ cases appear to be locally thermodynamically stable, however~\cite{Benincasa:2009be}. The specific heat at finite temperature scales with $T$ as
\beq
c_V = T\frac{\partial s}{\partial T} \sim \frac{T^\frac{2q}{z}}{d} \,.
\eeq
A linear scaling, as in a Fermi liquid, occurs when $z=2 q$, while a scaling $T^q$, as in a quantum Bose liquid, occurs when $z=2$~\cite{Lee:2010uy}.

\section{Green's Functions at Low Frequency}
\label{greens}

Zero sound should appear at zero temperature as a pole in the Fourier-transformed retarded two-point function of the density operator, $G_R^{tt}\left(\o,k\right)$ \cite{Karch:2008fa}. In what follows we will also compute the density-current and current-current retarded two-point functions $G_R^{tx}(\o,k)$ and $G_R^{xx}(\o,k)$. The latter will give us the AC conductivity.

In gauge-gravity duality the on-shell action is the field theory generating functional \cite{Gubser:1998bc,Witten:1998qj}, hence we need to take two functional derivatives of the on-shell bulk action, which in turn means we need to solve the linearized equations of motion for the fields dual to the density and current operators, namely the $U(1)$ gauge fields $A_t$ and $A_x$ on the probe D-branes. The holographic zero sound pole will appear as a quasi-normal frequency of these fields. We thus consider fluctuations $a_{\mu}$ of the gauge fields,
\beq
A_{\mu}(r) \rightarrow A_{\mu}(r) + a_{\mu}(r,t,x) \,,
\eeq
where because the field theory is invariant under rotations, we only need to consider fluctuations $a_t$ and $a_x$ with $r$, $t$, and $x$ dependence. We work in a gauge where $a_r = 0$. To quadratic order in the fluctuations, the action is
\beq
S_{a^2} = \frac{\N}{2}\!\int\!dr \, dt \, dx \, g_{xx}^{q/2} \left [ \frac{g_{rr} \tilde{f}_{tx}^2 - |g_{tt}|\tilde{a}_x^{'2}}{g_{xx}\sqrt{|g_{tt}| g_{rr} - \tilde{A}_t^{'2}}} + \frac{|g_{tt}| g_{rr} \tilde{a}_t^{'2}}{\left(|g_{tt}| g_{rr} - \tilde{A}_t^{'2} \right)^{3/2}} \right] ,
\eeq
where $f_{tx} = \partial_t a_x - \partial_x a_t$. Performing a Fourier transform,
\beq
a_{\mu}(r,t,x) = \int\!\frac{d\omega \, dq}{(2\pi)^2} \, e^{-i\omega t + i k x} \, a_{\mu}(r,\omega,k) \,,
\eeq
the linearized equations of motion become
\begin{align}
\partial_r \left [ \frac{g_{xx}^{q/2} |g_{tt}| g_{rr} a_t'}{\left(|g_{tt}| g_{rr} - \tilde{A}_t^{'2} \right)^{3/2}} \right] - \frac{g_{xx}^{q/2-1} g_{rr}}{\sqrt{|g_{tt}| g_{rr} - \tilde{A}_t^{'2}}} 
\left(k^2 a_t + \omega k a_x \right) &= 0 \,, \\
\partial_r \left [ \frac{g_{xx}^{q/2 - 1} |g_{tt}|a_x'}{\sqrt{|g_{tt}| g_{rr} - \tilde{A}_t^{'2} }} \right] + \frac{g_{xx}^{q/2-1} g_{rr}}{\sqrt{|g_{tt}| g_{rr} - \tilde{A}_t^{'2}}} 
\left(\omega^2 a_x + \omega k a_t\right) &= 0 \,.
\end{align}
Additionally, we find a constraint ($a_r$'s equation of motion, written in $a_r = 0$ gauge)
\beq
g_{rr} g_{xx}\omega a_t' + \left( |g_{tt}| g_{rr} - \tilde{A}_t^{'2} \right) k a_x' = 0 \,.
\eeq
If we solve the constraint equation for $a_x'$ and plug into the $a_x$ equation of motion, we find the $a_t$ equation of motion. We thus need only solve the constraint equation and $a_t$ equation. We will work with the gauge-invariant electric field (in Fourier space)
\beq
E(r,\omega,k) = \omega a_x + k a_t \,.
\eeq
The constraint equation and definition of $E$ give us
\beq
a_t' = \frac{u(r)^2 k}{u(r)^2 k^2 - \omega^2}E', \qquad 
a_x' = -\frac{\omega}{u(r)^2 k^2 - \omega^2}E' \,,
\eeq
where
\beq
u(r) = \sqrt{\frac{|g_{tt}| g_{rr} - \tilde{A}_t^{'2}}{g_{rr}g_{xx}}} = \sqrt{\frac{|g_{tt}|}{g_{xx}}}
\left(1 + d^2 g_{xx}^{-q}\right)^{-1/2} = r^{-(z-1)}\left(1 + d^2 r^{2q} \right)^{-1/2} \,.
\eeq
Plugging into the $a_t$ equation, we find (after some algebra) an equation for $E$,
\beq
\label{eq:Eeom}
E'' + \left [ \partial_r \ln \left(  \frac{g_{xx}^{(q-3)/2} g_{rr}^{-1/2}|g_{tt}|}{u
\left(k^2 u^2 - \omega^2\right)} \right) \right] E' - \frac{g_{rr}}{|g_{tt}|} 
\left(k^2 u^2 - \omega^2 \right) E = 0 \,.
\eeq

In terms of $E$ the quadratic action is 
\beq
S_{a^2} = \frac{\N}{2}\!\int\!dr \, d\omega \, dk \,\frac{g_{xx}^{(q-3)/2} g_{rr}^{1/2}}{u} \left [ \tilde{E}^2 +\frac{|g_{tt}|}{g_{rr}(u^2 k^2 - \omega^2)} \tilde{E}^{'\,2} \right] .
\eeq
Introducing a cutoff at $r=\epsilon$ and integrating by parts, we obtain in the $\epsilon\to 0$ limit
\beq
\label{eq:Eonshellaction}
S_{a^2} = -\frac{\N}{2}\!\int\!d\omega \, dk \frac{\epsilon^{z-q+1}}{k^2} \tilde{E}(\epsilon) \tilde{E}'(\epsilon) \,.
\eeq

We now need to solve the equation of motion with an ingoing boundary condition imposed at the $r\ra\infty$ ``horizon''~\cite{Son:2002sd,Herzog:2002pc,vanRees:2009rw}, insert the solution into $S_{a^2}$, and then functionally differentiate to obtain the retarded correlators, for example
\beq
G^{tt}_R(\omega,k) = \frac{\delta^2}{\delta a_t(\epsilon)^2}S_{a^2} =  \left(\frac{\delta E(\epsilon)}{\delta a_t(\epsilon)}\right)^2 \frac{\delta^2}{\delta E(\epsilon)^2} S_{a^2} \,.
\eeq
Defining
\beq
\Pi(\omega,k) \equiv \frac{\delta^2}{\delta E(\epsilon)^2} S_{a^2} \,,
\eeq
we get
\beq
\label{eq:pigrelation}
G^{tt}_R(\omega,k) = k^2\,\Pi(\omega,k)\,, \quad G^{tx}_R(\omega,k) = \omega\,k\,\Pi(\omega,k)\,, \quad G^{xx}_R(\omega,k) = \omega^2\,\Pi(\omega,k)\,.
\eeq
Our goal is to obtain (the low-frequency and low-momentum form of) the variational derivative $\Pi(\omega,k)$, which determines the retarded two-point functions that we want.

We have not been able to solve eq.~\eqref{eq:Eeom} exactly for all values of $\omega$ and $k$. A numerical solution is of course possible, however we can obtain the low-frequency behavior of $\Pi(\omega,k)$ by solving eq.~\eqref{eq:Eeom} in two different limits and then matching the two solutions in a regime where the limits overlap, following ref.~\cite{Karch:2008fa}. Specifically, we first obtain a solution by taking $r$ large and then taking the small frequency and momentum limit, which means $\omega r^z \ll 1$ and $k r \ll 1$ with $\omega k^{-z}$ fixed. We then repeat the process, now taking the small frequency and momentum limit first followed by taking $r$ large.

Taking $r \rightarrow \infty$, the equation for $E$ becomes
\beq\label{eq:horizoneq}
E'' + \frac{1}{r} \left [ 2 - (z-1) \right] E' + \omega^2 r^{2(z-1)} E = 0 \,.
\eeq
The solution we want involves a Hankel function of the first kind,
\beq
E = C\left(\frac{\omega r^z}{2z}\right)^{\frac{1}{2}-\frac{1}{z}}H^{(1)}_{\frac{1}{2}-\frac{1}{z}}\!
\left(\frac{\omega r^z}{z}\right) ,
\eeq
which for large $r$ describes an in-going wave: $E \sim \, r^{-1} \, e^{i \omega r^z/z}$ as $r \rightarrow \infty$. Now we want to take frequency and momentum small, which means in particular $\omega r^z \ll1$, since the solution above does not depend on $k$. The form of $E$ in this limit depends on the value of $z$. When $z \neq 2$, $E$ approaches
\beq\label{eq:largeuznoteq21}
E \approx C\,\Gamma\!\left(\frac{1}{z} + \frac{1}{2}\right)^{-1}\left(1 - i\tan\frac{\pi}{z} \right) - 
i\frac{C}{\pi}\,\Gamma\!\left(\frac{1}{z} - \frac{1}{2}\right)
\left(\frac{\omega}{2z}\right)^{1 - \frac{2}{z}}r^{z-2} \,.
\eeq
When $z=2$, the $\omega r^2 \ll 1$ limit of $E$ involves a logarithm
\beq
\label{eq:largeuzeq21}
E \approx C + C \, \frac{2i}{\pi} \left[\log \left( \omega r^2 \right)-\log 4+ \gamma \right],
\eeq
where $\gamma$ is the Euler-Mascheroni number.

Next, we start by taking $\omega r^z \ll 1$ and $k r \ll 1$ with $\omega k^{-z}$ fixed. The equation for $E$ becomes
\beq\label{eq:boundaryeq}
E'' + \left[ \frac{2-q-2(z-1)}{r} - \frac{u'}{u} \frac{3k^2 u^2 - \omega^2}{k^2 u^2 -\omega^2} \right] E' = 0 \,.
\eeq
The solution is
\begin{align}\label{eq:smallr}
E = C_1 + C_2\Bigg\{
k^2\frac{r^{q-z}}{q-z}\,
&{_2F_1}\!\left[\frac{q-z}{2q},\frac{3}{2};1+\frac{q-z}{2q};-d^2 r^{2q}\right] \notag \\ 
-\omega^2\frac{r^{-2+q+z}}{-2+q+z}\,
&{_2F_1}\!\left[\frac{-2+q+z}{2q},\frac{1}{2};1+\frac{-2+q+z}{2q};-d^2r^{2q}\right]
\Bigg\} \,, 
\end{align}
where $C_1$ and $C_2$ are constants. We can match to eqs.~\eqref{eq:largeuznoteq21} and~\eqref{eq:largeuzeq21} by taking $r$ large. The form of $E$ in this limit also depends on the value of $z$. When $z\neq 2$, the large $r$ limit of $E$ is ($B(x,y)$ is the Beta function) 
\beq
\label{eq:largeuznoteq22}
E \approx C_1  +  \frac{C_2}{d}\left[\frac{z k^2 d^{\frac{z}{q}}}{2q^2}B\!\left(\frac{1}{2} - \frac{z}{2q}, \frac{z}{2q}\right)- \frac{\omega^2 d^{\frac{2-z}{q}} }{2q}B\!\left(\frac{1}{2} + \frac{z-2}{2q},\frac{2-z}{2q}\right)  +  \frac{\omega^2 r^{z-2}}{(2-z)} \right] \,,
\eeq
which, as a function of $r$, consists of a constant and an $r^{z-2}$ term, which indeed match the terms in eq.~\eqref{eq:largeuznoteq21}. When $z=2$, the large $r$ limit of $E$ changes only in the $\omega^2$ term, 
\beq
\label{eq:largeuzeq22}
E \approx C_1 + \frac{C_2}{d}\left[\frac{k^2 d^{2/q}}{q^2 }B\!\left(\frac{1}{2} - \frac{1}{q},\frac{1}{q}\right) - \frac{\omega^2}{q}  \log \left(2 d r^q \right)\right] \,,
\eeq
which again consists of a constant and a logarithm, as in eq.~(\ref{eq:largeuzeq21}). To do the matching, we rewrite the logarithmic term as
\beq
\log\left(2 d r ^q\right) = \log \left( \frac{2 d}{\omega^{q/2}}\left(\omega r^2\right)^{q/2}\right) 
= \log \left(\frac{2 d}{\omega^{q/2}}\right) + \frac{q}{2} \log\left(\omega r^2\right) \,.
\eeq

The boundary behavior of the solutions is given by the small $r$ expansion of the solution in eq.~\eqref{eq:smallr}: $E\simeq C_1+ k^2 C_2 r^{q-z}/(q-z)$. When $z<q$, we have that to leading order $E(\epsilon) \simeq  C_1$, so
\beq
\Pi(\omega,k) =\lim_{\epsilon\to 0} \frac{\delta^2}{\delta E(\epsilon)^2} S_{a^2} = \left.\frac{\delta^2}{\delta C_1^2} S_{a^2}\right|_{\epsilon\to 0}\,,
\eeq
while to leading order $E'(\epsilon) \simeq k^2\epsilon^{q-z-1}C_2$, so\footnote{When $z\geq q$, $C_2$ becomes the source in the generating functional, and, as mentioned in section~\ref{thermo}, $S_{a^2}$ has new divergences as $\epsilon \rightarrow 0$. To make $S_{a^2}$ finite we add a boundary term, which for $z>q$ is
\beq
S_{\epsilon} = \frac{\N}{2} \int d\omega d\k \frac{q-z}{k^2}\sqrt{-\gamma}\, \gamma^{tt}\gamma^{xx} \tilde{E}^2(\epsilon) = \frac{\N}{2} \int d\omega dk \frac{q-z}{k^2} \epsilon^{z-q} \tilde{E}^2(\epsilon),
\eeq
and for $z=q$, where $E(\epsilon)\simeq C_1+C_2 k^2 \log \epsilon$, is
\beq
S_{\epsilon}=\frac{\N}{2}\int d\omega d\k \frac{1}{k^2\log\epsilon}\sqrt{-\gamma}\,\gamma^{tt}\gamma^{xx} \tilde{E}^2(\epsilon)=\frac{\N}{2}\int d\omega d\k \frac{1}{k^2\log\epsilon}  \tilde{E}^2(\epsilon),
\eeq
where $\gamma_{\mu\nu}$ is the induced metric at $r=\epsilon$. Adding such a term is equivalent to a Legendre transformation, which makes $C_1$ the source again. The result for $S_{a^2}$ is then identical to eq.~\eqref{eq:onshellsa2}, except with opposite sign. The boundary term does not change $E$'s equation of motion, so the solutions and matching are the same, while the overall sign is not crucial for our physical arguments. The $z\geq q$ cases are thus identical (up to a sign) to the $z < q$ cases, so we can freely take $z>q$ in our results.}
\beq
\label{eq:onshellsa2}
S_{a^2} = -\frac{\N}{2}\!\int\!d\omega \, dk \frac{\epsilon^{z-q+1}}{k^2} \tilde{E}(\epsilon) \tilde{E}'(\epsilon) = -\frac{\N}{2} (2\pi\alpha')^2 \!\int\!d\omega \, dk \, C_1 C_2 \,.
\eeq
To find $\Pi(\omega,k)$, we need to find $C_2$ in terms of $C_1$, which can be done by matching the constant term and the coefficient of the $r^{z-2}$ term in eqs.~(\ref{eq:largeuznoteq21}) and~(\ref{eq:largeuznoteq22}). More specifically, the matching gives us two equations involving $C$, $C_1$, and $C_2$, and we use one equation to eliminate $C$ so that the remaining equation gives us $C_2$ in terms of $C_1$. When $z=2$, we match the constant term and the coefficient of the logarithmic term $\log(\omega r^2)$ in eqs.~(\ref{eq:largeuzeq21}) and~(\ref{eq:largeuzeq22}). After that we immediately obtain $\Pi(\omega,k)$.

Assembling all the ingredients, we have our main result, the low-frequency and low-momentum form of $\Pi(\omega,k)$:
\beq
\label{eq:piresult}
\Pi(\omega,k)\propto \frac{\N}{\alpha_1 \, k^2 - \alpha_2 \, \omega^2- \alpha_3 \, G_0(\omega)} \,,
\eeq
where
\beq
G_0(\omega) = 
\begin{cases}
\omega^{1+2/z} & z\neq2 \, \\
\omega^2\log\left({\alpha\omega^2}\right) & z=2
\end{cases} \,,
\eeq
and $\alpha_1$, $\alpha_2$ and $\alpha_3$ are constants. For $z\neq2$ they are
\begin{align}
\alpha_1 &= \frac{2-z}{\pi\left(2z\right)^{1-2/z}}\,\Gamma\!\left(\frac{1}{z}-\frac{1}{2}\right) \Gamma\!\left(\frac{1}{z}+\frac{1}{2}\right)\frac{z \, d^{\frac{z}{q}}}{2q^2}\,B\!\left(\frac{1}{2} - \frac{z}{2q},\frac{z}{2q}\right) \,, \notag \\
\alpha_2 &= \frac{2-z}{\pi\left(2z\right)^{1-2/z}}\,\Gamma\!\left(\frac{1}{z}-\frac{1}{2}\right) \Gamma\!\left(\frac{1}{z}+\frac{1}{2}\right)\frac{d^{\frac{2-z}{q}}}{2q}B\!\left(\frac{1}{2} + \frac{z-2}{2q},\frac{2-z}{2q}\right) \,, \notag \\
\alpha_3 & = i +\tan\frac{\pi}{z} \,,
\end{align}
where for later reference we note that $\alpha_1$ and $\alpha_2$ are real while $\alpha_3$ is complex. For $z=2$ the constants are
\beq
\label{eq:z2coeffs}
\alpha_1 = \frac{4 \, d^{2/q}}{\pi} \, B\!\left(\frac{1}{2} - \frac{1}{q},\frac{1}{q}\right)\,, \quad \alpha_2 = i \,, \quad \alpha_3 = -\frac{1}{\pi} \,,
\eeq
and another (dimensionful) constant appears inside the logarithm in $G_0(\omega)$,
\beq
\alpha = \frac{e^{2\gamma}}{16} \left( 2d\right)^{-4/q} \,.
\eeq

In the denominator of $\Pi(\omega,k)$, when $z<2$ the $\omega^2$ term is larger than the $G_0(\omega)$ term, whereas when $z>2$ the $G_0(\omega)$ term is larger. The dominant term dictates the low-frequency behavior of the retarded Green's functions, and hence of the holographic zero sound and AC conductivity, as we will show in detail in what follows.

Notice that $z=2$ is a special value regardless of the number of spatial dimensions $q$, which is similar to what is observed for the worldvolume scalar that describes the embedding of the D-brane (without a magnetic field)~\cite{Jensen:2010ga}. To see more clearly how this happens in the bulk, let us consider the near boundary limit $r\to 0$ of eq.~\eqref{eq:boundaryeq}:
\beq
E''+\frac{1+z-q}{r} E' =0\,.
\eeq
We can compare this to the equation near the horizon, eq.~\eqref{eq:horizoneq}, when $\omega\to 0$:
\beq
E''+\frac{3-z}{r} E' =0\,.
\eeq
The near-horizon equation looks like the near-boundary equation with an effective number of spatial dimensions $q\to \tilde{q}=2(z-1)$. This suggests that the finite density theory flows towards an IR fixed point that depends only on the value of $z$, much like the holographic system of ref.~\cite{Faulkner:2009wj}, but restricted to the probe D-brane sector. If we also identify $\tilde{q}$ with the effective IR dimension of the density operator, then the operator $(J^t)^2$ becomes marginal when $\tilde{q}=z$, which happens precisely when $z=2$. More explicitly, when $z<2$ and $\tilde{q}<z$, $(J^t)^2$ is relevant, while when $z>2$ and $\tilde{q}>z$, $(J^t)^2$ is irrelevant, so $z=2$ would separate two different kinds of IR effective theories. Similarly to the logarithmic UV divergences we encountered in sec.~\ref{thermo}, the appearance of the logarithm in $G_0(\omega)$, which breaks scale invariance, may be due to $(J^t)^2$ becoming marginal at $z=2$ in the IR effective theory.

\section{Holographic Zero Sound}
\label{zero}

The denominator of $\Pi(\omega,k)$ vanishes -- and hence the retarded Green's functions have a pole -- whenever
\beq
\label{eq:ksol}
k(\omega) = \pm \frac{1}{\sqrt{\alpha_1}}\sqrt{\alpha_2 \, \omega^2 +\alpha_3 \, G_0(\omega)} \,.
\eeq
This determines the dispersion relation of the holographic zero sound mode: we need only expand the right-hand side of eq.~\eqref{eq:ksol} and then invert to find $\omega(k)$. As mentioned above, the value of $z$ determines whether the $\omega^2$ term or $G_0(\omega)$ term dominates at low frequencies, so we will consider two cases in turn.

\subsection{$1\leq z < 2$}

For $1 \leq z < 2$, the $\omega^2$ term under the square root has larger norm, so we expand $k(\omega)$ as
\bea
k(\omega) & = & \pm \, \omega \, \sqrt{\frac{\alpha_2}{\alpha_1}} \, \sqrt{1+\frac{\alpha_3}{\alpha_2} \, \omega^{-1+2/z}} \nonumber \\ & = & \pm \, \omega \, \sqrt{\frac{\alpha_2}{\alpha_1}} \, \left[1 +  \frac{\alpha_3}{2\alpha_2} \, \omega^{-1+2/z} + O\left( \omega^{-2+4/z} \right)\right] \nonumber ,
\eea
and then invert to find
\beq\label{eq:sounddisper}
\omega(k) = \pm \, k \, \sqrt{\frac{\alpha_1}{\alpha_2}} - \frac{\alpha_3}{2 \alpha_2} \left( \frac{\alpha_1}{\alpha_2} \right)^{1/z}k^{2/z} + O \left( k^{-1+4/z}\right).
\eeq
The mode behaves as a quasiparticle since the imaginary part goes as $k^{2/z}$, which is smaller than the real part at low momentum. As a check, if we set $z=1$ we recover the result of ref.~\cite{Karch:2008fa},
\beq
\omega(k) = \pm \frac{k}{\sqrt{q}} - i \frac{d^{-\frac{1}{q}}\Gamma\!\left(\frac{1}{2}\right)}{\Gamma\!\left(\frac{1}{2} - \frac{1}{2q}\right)\Gamma\!\left(\frac{1}{2q}\right)} \, k^2 + O\left(k^3\right).
\eeq

The speed of the holographic zero sound, $v_0$, is given by
\beq
\label{eq:a1overa2}
v_0^2=\frac{\alpha_1}{\alpha_2} = \frac{z}{q} \, d^{2\left(z-1\right)/q} \, 
\frac{\Gamma\!\left(\frac{z}{2q}\right)\Gamma\!\left(\frac{1}{2}-\frac{z}{2q}\right)}
{\Gamma\!\left(\frac{1}{2} + \frac{z-2}{2q}\right)\Gamma\!\left(\frac{2-z}{2q}\right)} \,.
\eeq
As promised in section~\ref{thermo}, when $z>1$ the speed of the holographic zero sound is dimensionful. In the relativistic case, $z=1$, the speed of holographic zero sound coincides with the speed of normal/first sound. For $1< z < 2$, the poles in $\Gamma\!\left(\frac{1}{2}-\frac{z}{2q}\right)$ and $\Gamma\!\left(\frac{2-z}{2q}\right)$ determine whether $v_0^2$ remains finite or goes to zero as $z \rightarrow 2$, with the details depending on the dimension $q$. When $q=2$, the $\Gamma$ functions (and hence their poles) immediately cancel and $v_0^2 = \frac{z}{2} d^{z-1}$, which is clearly finite as $z \rightarrow 2$. For any $q>2$, $\Gamma\!\left(\frac{1}{2}-\frac{z}{2q}\right)$ remains finite but $\Gamma\!\left(\frac{2-z}{2q}\right)$ has a pole as $z \rightarrow 2$, so $v_0^2$ goes to zero from above.

\subsection{$z\geq2$}

We consider first $z>2$. Now the $G_0(\omega)$ term under the square root in eq. (\ref{eq:ksol}) has larger norm, so we expand $k(\omega)$ as
\bea
k(\omega) & = & \pm\left( \frac{\alpha_3}{\alpha_1}\right)^\frac{1}{2}\omega^{\frac{z+2}{2z}}\left( 1 + \frac{\alpha_2}{\alpha_3}\omega^{1-\frac{2}{z}} \right)^\frac{1}{2} \nonumber \\ 
& = & \pm\left( \frac{\alpha_3}{\alpha_1}\right)^\frac{1}{2}\omega^{\frac{z+2}{2z}}\left[1 + \frac{\alpha_2}{\alpha_3}\omega^{1-\frac{2}{z}}  + O\left(\omega^{2-\frac{4}{z}}\right)\right] \,,
\eea
and invert to find
\beq
\omega(k) = \left( \frac{\alpha_1}{i \alpha_3}\right)^{\frac{z}{z+2}}k^{\frac{2z}{z+2}} + \frac{\alpha_2}{\alpha_3} \frac{z}{z+2} \left( \frac{\alpha_1}{\alpha_3}\right)^{\frac{2(z-1)}{z+2}}k^{\frac{4(z-1)}{z+2}}+O\left(k^{\frac{2(3z-4)}{z+2}}\right).
\eeq
Note that the leading term has a coefficient that is \textit{complex}. In particular, the real and imaginary parts are of the same order, hence the excitation is not a quasi-particle.

Finally, when $z=2$ we have
\beq
k = \pm\frac{\omega}{ \sqrt{\alpha_1}} \, \sqrt{\alpha_2 + \alpha_3 \log \left( \alpha \, \omega^2 \right)} .
\eeq
Expanding for small $\omega$ we find
\beq
k(\omega) = \pm\frac{\omega}{\sqrt{\alpha_1}} \, \sqrt{\alpha_3 \log \left(\alpha \, \omega^2\right)} - \frac{\omega}{\sqrt{\alpha_1}} \frac{\alpha_2}{2} \left( \alpha_3 \log\left(\alpha \omega^2\right)\right)^{-\frac{1}{2}} + O\left(\omega \log^{-\frac{3}{2}}\!\left(\alpha\omega^2\right)\right).
\eeq
Although this mode is gapless, the dispersion relation differs from the holographic zero sound mode by logarithmic factors.

\section{AC Conductivity}
\label{cond}

In this section we provide some physical intuition for the change in behavior of the holographic zero sound mode as a function of $z$ by studying the low-frequency AC conductivity $\sigma(\omega)$, defined as
\beq
\label{eq:conddef}
\sigma(\omega) \equiv -\frac{i}{\omega} \, G_R^{xx}(\omega,k=0)\,.
\eeq
Recalling that the imaginary part of the retarded Green's function is proportional to the spectral function of $J^x$, we see that the real part of the conductivity provides a measure of the density of states that couple to $J^x$.

To obtain $\sigma(\omega)$ for our system, we first set $k=0$ in our result for $\Pi(\omega,k)$ in eq.~\eqref{eq:piresult} and then plug into eq.~\eqref{eq:pigrelation}, which gives
\beq
G^{xx}_R(\omega,k=0) \propto  -\frac{\N \omega^2}{\alpha_2 \, \omega^2+ \alpha_3 \, G_0(\omega)}
\xrightarrow{\omega \rightarrow 0}\,-\N
\begin{cases}
\alpha_2^{-1} & z<2 \\
\alpha_3^{-1}\left(\log\left( \alpha \, \omega^2 \right)\right)^{-1} & z=2\\
\alpha_3^{-1}\omega^{1-2/z} & z>2
\end{cases}\,,
\eeq
and then obtain the AC conductivity at small frequency from eq.~\eqref{eq:conddef}
\beq
\sigma(\omega)=-\frac{i}{\omega} \, G^{xx}_R(\omega,k=0)\xrightarrow{\omega \rightarrow 0}\,\N
\begin{cases}
i\alpha_2^{-1}\omega^{-1} & z<2 \\
i\alpha_3^{-1}\left(\omega \log( \alpha \omega^2)\right)^{-1} & z=2\\
i\alpha_3^{-1}\omega^{-2/z} & z>2
\end{cases}\,.
\eeq
Our result for $\sigma(\omega)$ not only generalizes that of ref.~\cite{Hartnoll:2009ns} to any number of spatial dimensions $q$, but is an independent confirmation of the result, since we derived it via a slightly different route from that of ref.~\cite{Hartnoll:2009ns}. 

Recall that when $z\neq2$, $\alpha_2$ is a real number while $\alpha_3$ is complex. When $z<2$, the conductivity is purely imaginary and has a simple pole at zero frequency. A Kramers-Kronig relation then implies that the real part of the conductivity, and hence the spectral function, consists only of a delta function at zero frequency. The system has no charged states at small frequencies. Notice that the $i\omega^{-1}$ behavior is expected in a system with free charge carriers (or with translation invariance), and indeed is formally identical to the high-frequency or ``collisionless'' limit of the standard Drude result. When $z>2$ the conductivity, and hence the spectral function, has a power-law dependence, and no clean holographic zero sound quasi-particle exists.

Something analogous happens in interacting systems with a Fermi surface \cite{Pines:1966,Negele:1988vy}. Here the charge density spectral function includes a continuum of particle-hole pairs at frequencies $\omega < v_F k$, where $v_F$ is the Fermi velocity. The zero sound gives a contribution at $\omega = v_0 k$. If $v_0>v_F$, then the zero sound is outside the continuum, and will produce a sharp quasi-particle peak in the spectral function. However, if $v_0<v_F$, the zero sound mode suffers Landau damping through scattering with particle-hole pairs, and hence will be reduced to a broad resonance with no good quasi-particle interpretation. A similar mechanism appears to be at work in the holographic models we have studied: whenever the spectral function is non-vanishing at small frequencies, the holographic zero sound mode becomes a broad resonance.

Notice that previous calculations of the DC conductivity from holographic probe brane systems yielded a finite result \cite{Karch:2007pd}, whereas the na\"ive $\omega \rightarrow 0$ limit of our result for the AC conductivity is infinite for any value of $z$. This is not a contradiction. The different results come from the fact that the $\omega \rightarrow 0$ and $T \rightarrow 0$ limits do not in general commute, or equivalently the $\omega/T \rightarrow 0$ and $\omega/T \rightarrow \infty$ limits are physically distinct \cite{Herzog:2007ij}. The calculation of ref.~\cite{Karch:2007pd} involved taking $\omega \rightarrow 0$ first, which is the so-called collision-dominated or hydrodynamic limit $\omega/T \rightarrow 0$, while our calculation involves taking $T \rightarrow 0$ first, which is the collisionless limit $\omega/T \rightarrow \infty$. The latter is the appropriate limit in which to see zero sound in Fermi liquid systems \cite{Pines:1966,Negele:1988vy}.

\section{Discussion and Conclusion}
\label{discussion}

We determined the low-frequency and low-momentum form of the density and current retarded Green's functions for massless charge carriers in the holographic model of strange metals proposed in ref.~\cite{Hartnoll:2009ns}. Our main physical results are that for $z<2$ the system's spectrum includes a quasi-particle -- the holographic zero sound -- whose dispersion relation we computed, whereas when $z>2$ the quasi-particle is ``washed out'' by a continuum of low-frequency states. In addition, we recovered the results of ref.~\cite{Hartnoll:2009ns} for the AC conductivity from this single contribution to the spectral function. We also speculated that the change of behavior at $z=2$ may be due to the existence of an infrared fixed point where the effective dimension of the charge density is $2(z-1)$, so that the operator $(J^t)^2$ becomes marginal at $z=2$.

Our $z<2$ results are similar to those discovered in another holographic system, (3+1)-dimensional gravity in an extremal AdS-Reissner-Nordstr\"om geometry~\cite{Edalati:2009bi,Edalati:2010pn}. The dual field theory is a strongly-coupled (2+1)-dimensional CFT at finite density and zero temperature. Notice that here the density is not in the probe sector. Holographic calculations reveal a sound excitation in the spectrum, with a dispersion relation that is almost the $T \rightarrow 0$ limit of the normal/first sound dispersion: the speed squared is still equal to the conformal value $1/2$, while the sound attention constant differs by about $10\%$ \cite{Edalati:2010pn}. The leading contribution to the low-frequency AC conductivity is a pole in the imaginary part \cite{Edalati:2009bi}.

The field theory states dual to extremal AdS-Reissner-Nordstr\"om and to our solutions share two important features. First, both have a nonzero entropy at zero temperature. Second, both appear to have an emergent scaling symmetry in the infrared, indicating some fixed point involving strongly-coupled degrees of freedom \cite{Faulkner:2009wj}. Notice however that holographic zero sound has been found in systems with zero entropy at zero temperature \cite{Kulaxizi:2008jx}, which means that a nonzero $T=0$ entropy is not a necessary condition for the existence of a holographic zero sound mode. The absence of a gap for charged excitations appears to be a necessary condition, however \cite{Kulaxizi:2008kv}.

An alternative holographic approach to modeling strange metals is to introduce a charged probe fermion into AdS-Reissner-Nordstr\"om \cite{Liu:2009dm,Cubrovic:2009ye,Faulkner:2009wj,Faulkner:2010da}. The probe fermion is dual to a charged fermionic operator whose two-point functions reveal a Fermi surface with either Fermi-liquid or non-Fermi-liquid properties, depending on the bulk fermion's mass and charge \cite{Liu:2009dm,Cubrovic:2009ye,Faulkner:2009wj}. Given that the charge density is not in the probe sector, the fermions can only influence the charge and current Green's functions via back-reaction or one-loop effects. A calculation of the latter reveals that the probe fermions' contribution to the resistivity can scale linearly with temperature~\cite{Faulkner:2010da}. An open question is whether back-reaction or one-loop effects produce a zero sound pole, which may be a genuine Landau zero sound, corresponding to fluctuations in the shape of the Fermi surface.

A related approach is to introduce a weak coupling between a Fermi liquid and a critical sector described by the near-horizon region of the extremal AdS-Reissner-Nordstr\"om black hole \cite{Faulkner:2010tq}. In this case the strange metal behavior comes from the critical sector's contribution to the fermion self-energy. A zero sound mode would receive corrections to its self-energy as well, modifying its dispersion relation. We have found indications that the near-horizon region of the probe D-branes also describes an IR critical theory, so we expect that an analysis similar to that of ref.~\cite{Faulkner:2010tq} could be possible here, and that for the zero sound it would produce results similar to ours.

Returning to the model of ref.~\cite{Hartnoll:2009ns}, many avenues remain open for future research. As emphasized in ref.~\cite{Hartnoll:2009ns}, a realistic analysis requires massive charge carriers, with a mass gap much larger than the temperature. In the bulk that means introducing a scalar field dual to the flavor mass operator on the probe D-brane worldvolume. In relativistic string theory systems, a special cancellation of metric factors occurs in the DBI action, allowing for an analytic solution not only for $A_t'(r)$ but also for the scalar \cite{Karch:2007br}. These analytic solutions then allow for an analytic calculation of the holographic zero sound dispersion relation \cite{Kulaxizi:2008kv}. The cancellation of warp factors does not occur for generic $z$, at least with a trivial dilaton.  Nevertheless, we expect the physics for $z>1$ to be similar to the $z=1$ case, at least whenever a holographic zero sound quasi-particle exists. For example, we expect the speed of the holographic zero sound to go to zero as the chemical potential approaches the value of the mass from above \cite{Kulaxizi:2008kv}.

In relativistic string theory settings, the holographic zero sound quasi-particle is absent at any finite temperature, apparently as a consequence of the probe limit \cite{Kim:2008bv}. In these systems, a finite temperature produces an energy density of order $N^2$, which is enormous compared to the probe flavor's order $N_f N$ energy density. The low-frequency dynamics is thus dominated by adjoint-sector physics, which induces charge diffusion. The spectral function grows linearly with frequency in this case, so the finite-temperature width of the holographic zero sound is big and the pole is smeared into a broad resonance. Presumably the same is true for any $z<2$. A natural question is what happens if we leave the probe limit. Does the holographic zero sound remain a sharp quasi-particle when $N_f$ is on the order of $N$? Notice also that a fully back-reacted calculation would unify several previous calculations, at least in systems with $z=1$: at small field strength, where the DBI action is well-approximated by a Maxwell action, the results must reduce to those of ref.~\cite{Edalati:2010pn} (at least in (2+1) dimensions), while in a probe limit the results must reduce to those of ref.~\cite{Karch:2008fa}. Notice that in both limits the speed of the sound mode was identical to the speed of normal/first sound.

Lastly, the model building proposed in ref.~\cite{Hartnoll:2009ns} is worth pursuing (see for example ref.~\cite{Lee:2010ii}). As noted there, some studies of the high-$T_c$ cuprate strange metals find that the low-frequency AC conductivity scales as $\omega^{-2/3}$, which suggests that any holographic model reproducing such scaling may not exhibit a zero sound pole. A holographic system that reproduces all measured strange metal properties will be more complicated than just probe D-branes in a Lifshitz geometry however, so the question of whether a holographic zero sound quasi-particle exists in these more complicated models should be studied on a case-by-case basis.

\section*{Acknowledgments}

We would like to thank A.~Andreev, S.~Hartnoll, K.~Jensen, A.~Karch, R.~Meyer, D.~T.~Son, D.~Tong, and J.~Zaanen for helpful conversations. We also thank M.~Ammon for collaboration during the early stages of this project. The work of C.H. is supported in part by the U.S. Department of Energy under Grant No. DE-FG02-96ER40956. The work of A.O'B. is supported in part by the Cluster of Excellence ``Origin and Structure of the Universe.'' The work of J.W. is supported in part by the ``Innovations- und Kooperationsprojekt C-13'' of the ``Schweizerische Universit\"atskonferenz SUK/CRUS'' and the Swiss National Science Foundation.

\bibliography{lifshitz_zero_sound_v2b}
\bibliographystyle{JHEP}

\end{document}